\documentclass[9pt,shortpaper,twoside,web]{ieeecolor}
\usepackage{generic}
\usepackage{cite}
\usepackage{amsmath,amssymb,amsfonts}
\usepackage{algorithmic}
\usepackage{graphicx}
\usepackage{textcomp}
\usepackage{multirow,multicol} 
\usepackage{array} 
\usepackage{bm}
\usepackage{url}
\usepackage{epstopdf}

\def\BibTeX{{\rm B\kern-.05em{\sc i\kern-.025em b}\kern-.08em
    T\kern-.1667em\lower.7ex\hbox{E}\kern-.125emX}}
\begin{document}
\title{Combating Ambiguity for Hash-code Learning in Medical Instance Retrieval}
\author{Jiansheng Fang, Huazhu Fu, Dan Zeng, Xiao Yan, Yuguang Yan, and Jiang Liu
\thanks{This work was supported in part by the Science and Technology Innovation Committee of Shenzhen City (20200925174052004 and JCYJ20200109140820699). }
\thanks{J.Fang is with the School of Computer Science and Technology, Harbin Institute of Technology, Harbin, China, and also with the Department of Computer Science and Engineering, Southern University of Science and Technology, Shenzhen, China, and also with CVTE Research, Guangzhou, China (e-mail: 11949039@mail.sustech.edu.cn).}
\thanks{H.Fu is with the Inception Institute of Artificial Intelligence, Abu Dhabi, United Arab Emirates (e-mail: hzfu@ieee.org).}
\thanks{D.Zeng and X.Yan are with the Department of Computer Science and Engineering, Southern University of Science and Technology, Shenzhen, China.}
\thanks{Y.Yan is with the Department of Mathematics, University of Hong Kong, Hong Kong, China.}
\thanks{J.Liu is with the Department of Computer Science and Engineering, Southern University of Science and Technology, Shenzhen, China, and also with the Cixi Institute of Biomedical Engineering, Chinese Academy of Sciences, Ningbo, China (e-mail: liuj@sustech.edu.cn).}
\thanks{Corresponding author: Jiang Liu.}
}

\maketitle

\begin{abstract}
When encountering a dubious diagnostic case, medical instance retrieval can help radiologists make evidence-based diagnoses by finding images containing instances similar to a query case from a large image database. The similarity between the query case and retrieved similar cases is determined by visual features extracted from pathologically abnormal regions. However, the manifestation of these regions often lacks specificity, i.e., different diseases can have the same manifestation, and different manifestations may occur at different stages of the same disease. To combat the manifestation ambiguity in medical instance retrieval, we propose a novel deep framework called Y-Net, encoding images into compact hash-codes generated from convolutional features by feature aggregation. Y-Net can learn highly discriminative convolutional features by unifying the pixel-wise segmentation loss and classification loss. The segmentation loss allows exploring subtle spatial differences for good spatial-discriminability while the classification loss utilizes class-aware semantic information for good semantic-separability. As a result, Y-Net can enhance the visual features in pathologically abnormal regions and suppress the disturbing of the background during model training, which could effectively embed discriminative features into the hash-codes in the retrieval stage. Extensive experiments on two medical image datasets demonstrate that Y-Net can alleviate the ambiguity of pathologically abnormal regions and its retrieval performance outperforms the state-of-the-art method by an average of $9.27\%$ on the returned list of $10$.
\end{abstract}


\section{Introduction}
Content-based image retrieval (CBIR) has been mostly tackled as the problem of instance-level image retrieval \cite{zheng2017sift} and has been a long-standing research topic in the computer vision society \cite{zhou2017recent}. When encountering a dubious diagnostic case, CBIR systems can help radiologists search for similar cases in their decision-making process. Instance-level image retrieval is to hunt for images with the same instance as a query image in a large image database \cite{xiao2020deeply}. The benefit of instance-level retrieval for medical image screening and diagnosing can be witnessed in an observer study \cite{li2003investigation}. Five participating radiologists were given the task of querying nodules, for which they were required to infer the likelihood of malignancy. The task was performed twice: once with the aid of the search engine and once not. The search engine returned $3$ instances of the most similar malignant images and $3$ instances of the most similar benign images to help these radiologists making the inference. The average performance of the five radiologists was shown to increase from $0.56$ to $0.63$ with the aid of similar nodules. 

The relevancy of instance-level retrieval is mainly grounded on the visual similarity of instances rather than the whole image \cite{zhan2018instance}, so the features of a region-wise instance residing in a retrieved image should be explored effectively. Recently, many existing works on instance-level retrieval typically extracted visual features by using convolutional neural networks (CNN) to prevent the visual features unique to an instance from drowning in the global image. Early works \cite{babenko2014neural,li2015feature} focused on replacing traditional hand-crafted descriptors with features from fully-connected layers. The second generation of works \cite {babenko2015aggregating,razavian2016visual} achieved significant gains by encoding the activations of convolutional layers as region-wise feature descriptors. Among CNN-based approaches of instance-level retrieval, deep hashing methods \cite{zhuang2016fast,conjeti2017deep} have arisen as a promising solution because of their efficient data storage and fast searching. 

\begin{figure}[htbp]
  \centering
  \includegraphics[width=0.95\linewidth]{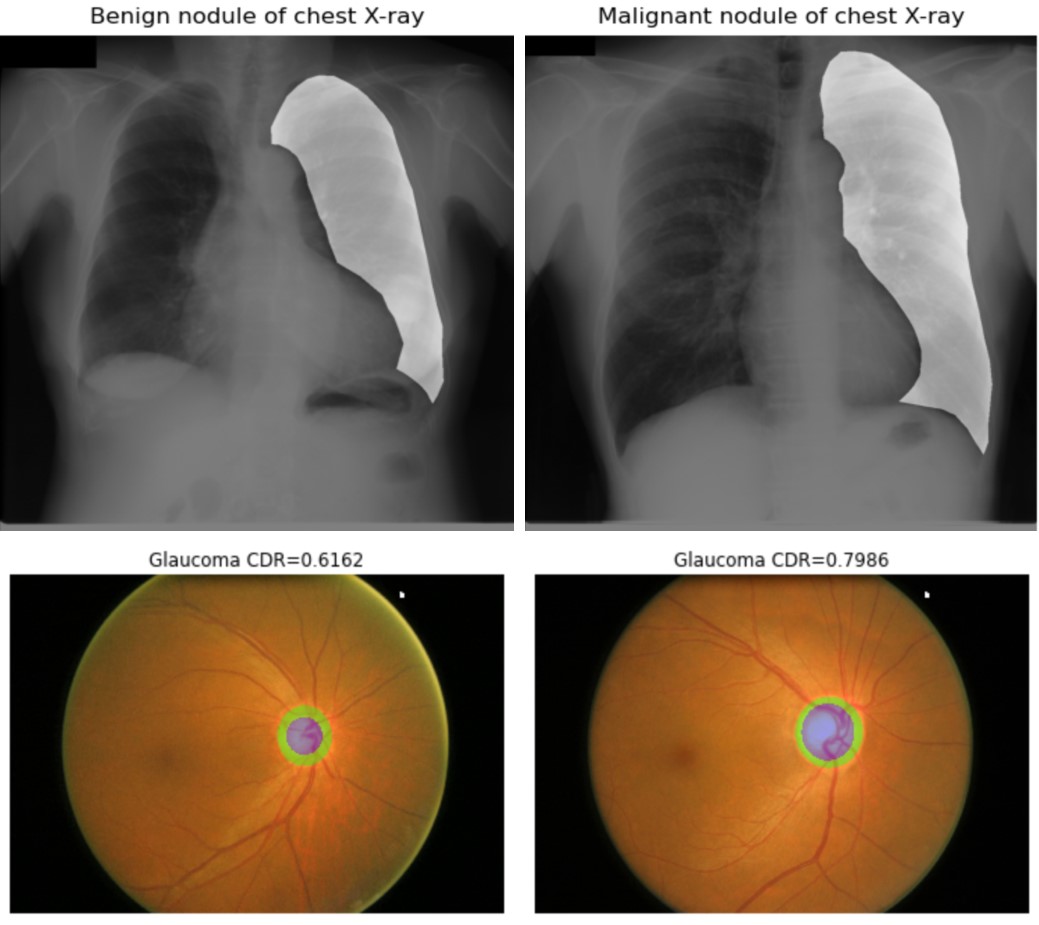}
  \caption{Illustration of ambiguity in medical image diagnosis. The upper row is two chest X-ray images labeled benign and malignant nodule. The down row is two glaucoma fundus images with a cup to disk ratio (CDR).}
  \label{fig1}
\end{figure}

Deep hashing methods can preserve the information of high-dimensional images by jointly learning image descriptors and hash-codes in an end-to-end framework \cite{do2019compact,fang2020attention,fang2021deep}. The image descriptors from fully-connected layers or convolutional layers are mapped into compact hash-codes for similarity comparison. Existing deep hash methods for instance-level retrieval have been shown to be effective and efficient \cite{wu2019deep,yang2020deep}. However, generating hash-codes in medical instance retrieval is challenging due to the manifestation ambiguity of pathologically abnormal regions. Such an issue plagues radiologists in the clinically routine screening and largely affects medical instance retrieval performance. It can be varied in two kinds: different diseases can have the same pathological abnormalities (\textbf{SPDD}), while different pathological manifestations may occur at different stages of the same disease (\textbf{DPSD}). As Fig. \ref{fig1} shows, 1) \textbf{SPDD} problem: it is difficult to interpret chest X-ray images and recognize the subtle difference between malignant and benign nodules, the lesion region of both images is on the left lung's upper lobe and has similar manifestations. However, the malignant image is diagnosed as lung cancer, and the benign image is pulmonary hematoma. Only professional radiologists can find the difference between benign and malignant nodules. 2) \textbf{DPSD} problem: cup to disk ratio (CDR), which is the ratio of cup diameter to disc diameter and often be employed as the main clue of glaucoma diagnose, varies at different stages.

The ambiguity of pathologically abnormal regions may prevent the assimilation of medical instance retrieval into an assistant tool for medico-decision \cite{loyman2020semi}. One solution is to provide ground-truth fine-grained labels to combat the ambiguity of pathologically abnormal regions. However, medical annotations remain highly dependant on manual expert feedback with high inter-observer variability \cite{faruque2015content}. Generally, medical image datasets can provide class labels for classification and pixel-wise masks for segmentation. Hence, a feasible solution is to effectively exploit the visual contents of pathologically abnormal regions based on class labels and pixel-wise masks \cite{li2018large}. Following this way, we present a novel deep framework, called Y-Net, to learn deep representations from image spaces by unifying segmentation and classification losses. During the training stage, the spatially subtle differences and class-aware semantic information of pathological regions are simultaneously learned into convolutional features. In the test stage, the learned convolutional features are aggregated into the hash-codes to preserve visual features unique to pathologically abnormal regions. 

There are two main motivations to present the Y-Net framework for alleviating the specificity shortage in medical instance retrieval. First, traditional deep hashing networks are to learn the global descriptor in an end-to-end way. They are prone to make the discriminative regions drown in the global descriptor. On the contrary, our Y-Net aims to explore the pixel-wise discriminative information by segmentation guidance, which pays more attention on the pathologically abnormal regions. Second, existing instance retrieval methods using local aggregation usually locate local regions in an unsupervised or weakly-supervised manner, which ignores the label information, while our Y-Net exploits class labels to locate the discriminative regions. The main contributions of this work are summarized as follows:
\begin{itemize}
    \item[1)] To combat ambiguity of pathologically abnormal regions in medical instance retrieval, we present a novel Y shape deep network, named Y-Net, encoding images into compact hash-codes. Our Y-Net can improve the differentiating ability of the hash-codes by exploiting the visual features unique to pathologically abnormal regions.
    \item[2)] Y-Net unifies classification and pixel-wise segmentation training to learn good semantic-separability and spatial-discriminability convolutional features. The segmentation branch learns subtle spatial differences to avoid the \textbf{SPDD} problem while the classification branch locates the discriminative regions by class-aware semantic information to overcome the \textbf{DPSD} problem. 
    \item[3)] Extensive experiments on two public medical datasets demonstrate that our proposed Y-Net can further improve the retrieval performance compared to the state-of-the-art instance retrieval methods. Our code and model have been released in \url{https://github.com/fjssharpsword/YNet}. 
\end{itemize}

The rest of this work is organized as follows: Section \uppercase\expandafter{\romannumeral2} discusses related works. Section \uppercase\expandafter{\romannumeral3} describes our methodology in detail. Section \uppercase\expandafter{\romannumeral4} extensively evaluates the proposed method on two medical images datasets. Section \uppercase\expandafter{\romannumeral5} gives concluding remarks.

\section{Related Works}
This section gives some related works that have contributed to instance-level retrieval and discuss current research progress of medical instance retrieval.

\subsection{Instance-level Retrieval}
Hashing methods can be divided into data-independent methods and data-dependent methods. The data-independent methods \cite{slaney2008locality,raginsky2009locality} learn hashing functions in a two-stage manner from hand-crafted features such, and the hash-codes learning procedure is independent of the image features, which may lead to sub-optimal performance. The data-dependent methods, also called learning-based hashing methods, can be further categorized into \cite{chen2018order}: (1) shallow learning-based hashing methods, like metric hashing forests \cite{conjeti2016metric}, and kernel sensitive hashing \cite{liu2016deep}; (2) deep learning-based hashing methods, like image inpainting-based compact hash-code learning \cite{ozturk2020image}, and deep hashing network \cite{zhu2016deep}. In contrast to the data-independent methods, they extract global features for hashing in an end-to-end manner. Early works \cite{jegou2008hamming,arandjelovic2011smooth} for instance-level retrieval rely on hand-crafted local descriptors such as SIFT \cite{lowe2004distinctive} and SURF \cite{bay2006surf}. Prior to deep learning, these works based on local features extraction, then aggregated into a global vector \cite{tolias2013aggregate,tolias2014orientation}.
The instances relevant to a query are discovered in the candidate images for similarity search by matching local descriptors. However, hand-crafted local features are vulnerable to non-rigid deformations and heavy viewpoint changes. Due to the promising performance in computer vision, CNN-based approaches have been introduced to instance-level retrieval. Instead, the global vector is extracted by a single forward-pass through a CNN, in which the extraction and aggregation steps are not separated. Existing deep hashing methods \cite{ozturk2020stacked,ozturktwo} can be grouped into this category using feature embedding tailor features from fully-connected layers for hash-codes generating. The representative methods include deep pairwise-supervised hashing (DPSH) \cite{li2015feature}, deep supervised hashing (DSH) \cite{wang2016deep}, and deep residual hashing (DRH) \cite{conjeti2017hashing}. Since convolutional features have been found to be reasonably discriminative \cite{kim2018regional}, recent CNN-based approaches have shifted to concentrate on feature aggregating rather than feature embedding. CNN-based approaches aggregating convolutional feature maps as global image representation can be roughly divided into two categories.

The first category is the works encoding the activations of a convolutional layer by weighted aggregation. These works' key idea is to assign different weights to different regions' activations in feature maps after global convolutional layers generate. SPoC \cite{babenko2015aggregating} showed that a simple spatial pooling on the convolutional layer outperformed fully-connected layers, and the power of this representation could be enhanced by applying the Gaussian center prior scheme to weight the contribution of the activations before aggregation. Following a similar idea, CroW \cite{kalantidis2016cross} proposed a non-parametric spatial- and channel-wise weighting method for focusing on salient regions. Unlike the spatial weighting scheme, class activation maps (CAMs) \cite{jimenez2017class} are employed for calculating semantic-aware weights of a convolutional feature map. Based on the bags of local convolutional features (BLCF) \cite{mohedano2016bags}, BLCF-SALGAN \cite{mohedano2018saliency} build an efficient image representation by saliency weighting. 

The second category is the works performing region analysis using convolutional features. Unlike the first category, this category first generates regions' convolutional features after region proposal, then aggregates them into global features. The representative work is regional maximum activation of convolutions (R-MAC) \cite{tolias2015particular}, which generated a set of regional vectors by performing spatial max-pooling within a particular region and aggregates features from several local regions into a single compact feature. Gordo \textit{et al.} \cite{gordo2017end} improved over the original R-MAC encoding by explicitly learning a region proposal network \cite{ren2015faster} and training in an end-to-end framework with a triplet loss. Laskar \textit{et al.} \cite{laskar2017context} used a saliency measure directly derived from the convolutional features to weigh the contribution of the regions of R-MAC before aggregation. Similar to R-MAC, Cao \textit{et al.} \cite{cao2016focus} proposed a method to derive a set of base regions directly from the convolutional layer, followed by a query adaptive re-ranking strategy. DeepVision \cite{salvador2016faster} extracted region-level features from the bounding boxes generated by the object detection framework. Regional attention \cite{kim2018regional} proposed a context-aware regional attention network that weighs an attentive score of a region considering global attentiveness. 

\subsection{Medical Instance Retrieval}
Recently, deep hashing methods using feature embedding on the fully-connected layer have also been widely proposed for medical instance retrieval, such as deep multiple instances hashing for tumor assessment \cite{conjeti2017deep}, deep residual hashing for chest X-ray images \cite{conjeti2017hashing}, order-sensitive deep hashing method for multi-morbidity medical image retrieval \cite{chen2018order}, deep disentangled momentum hashing for Neuroimage Search \cite{yang2020deep}, etc. Although the prior works have facilitated medical instance retrieval's prosperity, pathologically abnormal regions' manifestation ambiguity is challenging for current deep hashing methods. Recent studies \cite{cao2016focus,jimenez2017class,kim2018regional} have shown that using feature aggregating on the convolutional features achieves promising performance in instance-level retrieval. Following this direction, our work improves the current deep hashing method to combat pathologically abnormal regions' ambiguity in medical instance retrieval. 

In this work, the improvement for the current deep hashing methods includes: 
\begin{itemize}
    \item Unlike the current deep hashing methods jointly learning image descriptors and hash-codes, our work first learns convolutional features from image spaces by supervised training, then aggregates them as hash-codes. The learned convolutional features and following generated hash-codes can effectively preserve the differentiating information of pathologically abnormal regions.
    \item Inspired to CAMs \cite{jimenez2017class} and R-MAC \cite{tolias2015particular}, we endow the class-aware information to the R-MAC descriptors by classification training. The R-MAC descriptors related to classes can enhance their differentiating ability and help to avoid the \textbf{SPDD} problem. 
    \item Motivated by regional attention \cite{kim2018regional}, we adopt feature pyramid networks (FPN) \cite{lin2017feature} to exploit multi-scale pathologically abnormal regions by pixel-wise segmentation training. The subtle differences are encoded into the convolutional features to overcome the \textbf{DPSD} problem. 
\end{itemize}
We detect if pathologically abnormal regions are presented in each image with classification training, and we locate pathologically abnormal regions using activations with the help of segmentation training. In the end, the convolutional features, having learned class-aware information and subtle spatial differences, are mapped into the hash-codes. Based on the class labels and pixel-wise masks, we argue that our work is a beneficial exploration to combat pathologically abnormal regions' ambiguity in medical instance retrieval.

\section{Methodology}
Our Y-Net aims to generate highly distinctive hash-codes from the learned convolutional features. The hash-codes should meet three requirements: ($a$) the query image should be encoded close to positive images with the same instance and far from negative images without the same instance in the hashing space; ($b$) the class-aware semantic information and subtle differences of pathologically abnormal regions should be effectively encoded in convolutional features; ($c$) The convolutional features should be effectively aggregated to the compact hash-codes to preserve the learned visual cues. This section will elaborate on our Y-Net, including the main branch, R-MAC branch, FPN branch, and the coupled loss function.

\subsection{Framework Overview}
\begin{figure*}[htbp]
  \centering
  \includegraphics[width=0.95\linewidth]{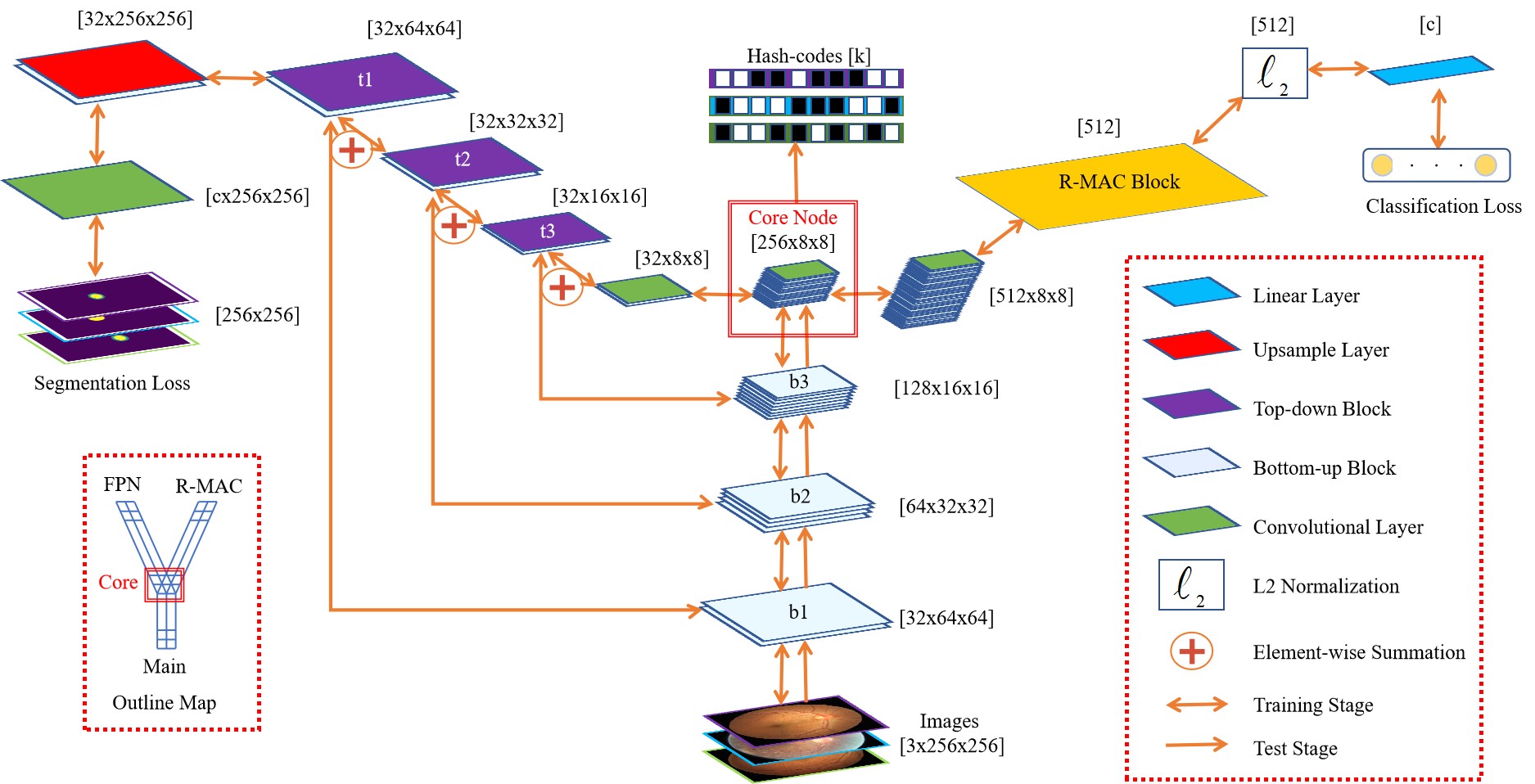}
  \caption{A novel deep framework for medical instance retrieval. We feed an image to the main branch, followed by the R-MAC branch (right) for classification training and the FPN branch (left) for segmentation training. The class-aware semantic information of pathological regions from the R-MAC branch and the spatially subtle differences of pathological regions from the FPN branch is effectively learned in the convolutional feature maps in the core node (red rectangle). The convolutional feature maps are mapped into the hash-codes via feature aggregating in the test stage. Our framework's shape is similar to a Y shape, including the main branch, the R-MAC branch, and the FPN branch, called Y-Net.}
  \label{fig2}
\end{figure*}
Each image is represented by an instance-invariant feature vector, i.e., hash-code. As shown in Fig.\ref{fig2}, we present a deep learning framework, called Y-net, to generate distinctive hash-codes from convolutional features. Our Y-Net contains three parts, main branch, R-MAC branch (right), and FPN branch (left). In the training stage (double arrow), we input an image into the main branch and feed-forward to the core node (red rectangle). The core node is a convolutional layer, followed by the R-MAC branch and the FPN branch. In the R-MAC branch, the classification loss minimizes intra-class distance and maximize inter-class distance. The inter-class separation can help avoid \textbf{SPDD} problem. But, to overcome the \textbf{DPSD} problem, the intra-class distance needs to be preserved but not minimized. The FPN branch can locate intra-class differences by pixel-wise segmentation training to balance the reduction of intra-class distance in the R-MAC branch. The core node learns the class-aware semantic information of pathological regions from the R-MAC branch for differentiating the same manifestation of different diseases. Simultaneously, the spatially subtle differences of pathological regions from the FPN branch are encoded into the core node to locate the same disease's subtle differences at different stages. After the core node absorbing the visual cues from the R-MAC branch and the FPN branch in the training stage, we can generate hash-codes from the learned core node by feature aggregation in the test stage (single arrow). 

\subsection{Main Branch}
The image encoding pipeline of the main branch is depicted as follows:
\begin{itemize}
    \item \textbf{Training Stage.} The input image $\bm{I}$ with a resolution $3\times 256\times 256$ is feed-forwarded into the main branch. The main branch computes a feature hierarchy consisting of a bottom-up block at three scales with a scaling step of $2$. At each scale, we use the feature activations output of bottom-up block \cite{he2016deep} to get a receptive field. The three bottom-up blocks are merged into the FPN branch by addition. In the core node of the main branch, the convolutional feature maps $\bm{X} \in \mathbb{R}^{C\times H\times W}$ can be arranged in a tensor of the size $C\times H\times W$, where $H$ and $W$ denote the height and width of each feature map, and $C$ denotes the number of feature maps (or channels) in the convolutional layer. In a convolutional layer, the activations at the same spatial location across all feature maps can be composed into a $C$-dimensional local descriptor for a certain image region. Compared to the activations of the fully-connected layer, the convolutional features retain the spatial information of local image descriptors and are essentially similar to the traditional hand-crafted local features \cite{long2014convnets,liu2015treasure}. The convolutional feature maps $\bm{X}$ are further feed-forwarded into the R-MAC branch and the FPN branch. Based on the classification and segmentation training, the semantic and spatial information of pathological regions is encoded into the convolutional feature maps in the feedback process.
     \item \textbf{Test Stage.} Based on the pre-trained Y-Net, an image with a resolution $3\times 236\times 256$ is feed-forwarded into the main branch and terminated in the core node. The core node is the conjunct point of a Y shape and is the core component in the framework of Y-Net. 
     We apply feature aggregation to generate a $k$-bits hash-code from the learned convolutional feature maps $\bm{X}$ with $C\times H\times W$ in the core node. Based on the existing works, the feature aggregation does not take part in the training and has been found to be more capable of preserving the discriminative information than the feature embedding. Considering the convolutional feature maps $\bm{X}$ with $C\times H\times W$ have learned the visual cues of pathological regions effectively, we convolute the size of $C\times H\times W$ into the size of $c\times h\times w$ without any weighting strategy. Then the three dimensions vectors further are squeezed into one dimension; its size equals the hash-code size of $k$-bits. Such a convolution process can aggregate feature maps of various sizes in three dimensions, such as $1\times 8\times 8$ and $128\times 1\times 1$. Lastly, we apply the hyperbolic tangent function to generate the value between $-1$ and $1$, following by signed as binary hash-code. At this step, we do not introduce any weighting strategy on feature aggregation because the convolutional feature maps have learned the visual cues of pathological regions effectively. 
\end{itemize}

\subsection{R-MAC Branch}
The R-Mac branch contains a convolutional layer using $3\times 3$ filters and followed by batch normalization \cite{ioffe2015batch}, then an R-MAC block generating a feature vector of length $512$. The feature vector is mapped into a linear layer after the $L_{2}$ normalization. The length of the linear layer is the number of classes. The R-MAC block generates a compact representation by aggregating multiple regions at different scales. By classification training, the highly activated regions can correspondingly respond to the semantic information of the belonging class. The pipeline of the R-MAC block is summarized as follows:
\begin{itemize}
    \item Based on a convolutional layer with $512\times 8\times 8$, we sample square regions with a region size, $R_{s}$, of a specific scale $s$ in a sliding window manner of $0.4$ overlap between neighbor windows, for all $s=0,...,S$. The region size at a specific scale can be calculated as:
        \begin{equation}
            \begin{aligned} 
            R_{s} = 2 \times \min(W_{r},H_{r}) / (s+2)
            \end{aligned}
            \label{eq1}, 
        \end{equation}
    where $W_{r}$ and $H_{r}$ are width and height of the feature map in the convolutional layer. In our Y-Net, with $W_{r}=8$ and $H_{r}=8$, we set $S=3$, then we totally get sample region of $14$.
    \item After sampling the regional feature maps, we perform a max-pooling for all regional feature maps of $14$. Each regional feature maps generate a feature vector with $512$, the same as the channel's size. Last, we aggregate all feature vector of sample regions in the whole image as a global feature vector with $512$ dimensions, named R-MAC descriptor used as a discriminative image representation.
\end{itemize}
In the pipeline of R-MAC, the local features from a certain convolutional layer are max-pooled across several multi-scale overlapping regions, obtained from a rigid grid covering the whole image, similar in spirit to spatial pyramids, producing a single feature vector per region. Then these region descriptors are sum-aggregated and $L_{2}-$ normalized into a global image representation. The discriminative global image representation is a compact vector whose size is independent of the size of the image and the number of regions. The region pooling is different from a spatial pyramid. The latter concatenates the region descriptors, while the former sum-aggregates them. Comparing the R-MAC descriptors of two images with a dot-product can then be interpreted as a many-to-many region matching.

\begin{figure}[htbp]
  \centering
  \includegraphics[width=0.95\linewidth]{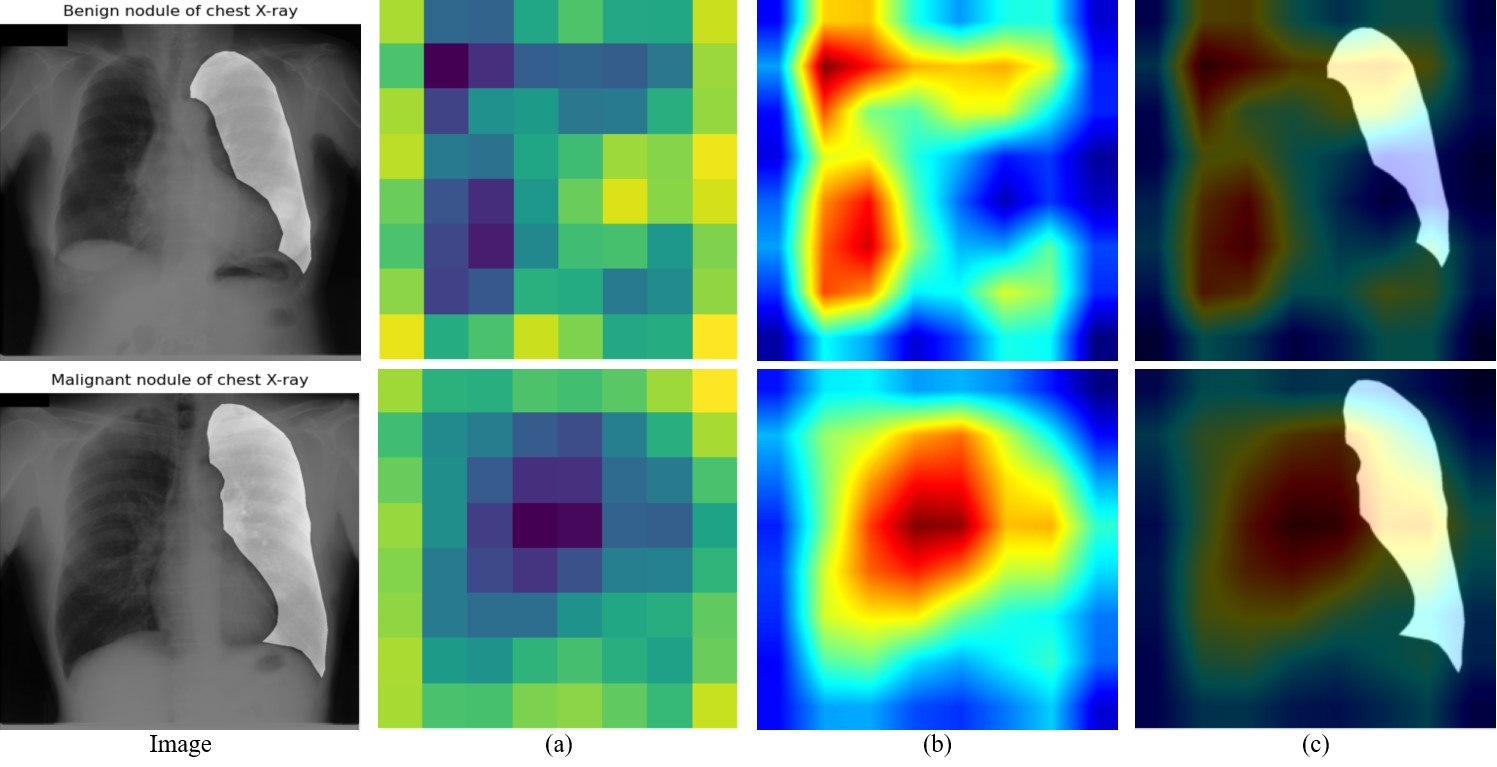}
  \caption{Feature maps of the R-MAC branch. We calculate the mean value along the channel axis of the convolutional features with $512\times 8\times 8$, then visualize the mean value: (a) the feature map with $8\times 8$, (b) the color map resized to the size of the mask image with $256\times 256$, and (c) the overlay map combined the color map with the mask image.}
  \label{fig3}
\end{figure}
R-MAC has been known for effective and efficient performance in image retrieval. Nonetheless, the main issue of R-MAC is that all sampled regions are equally treated without considering their varying importance. When aggregating their regional feature vectors, all regions construct their equal attentiveness to the last R-MAC descriptor. To overcome this problem, we integrate R-MAC in our Y-Net for supervised training to avoid the class-agnostic problem of the descriptors in R-MAC. We argue that the convolutional layer activations before the R-MAC block can respond to the semantic information during classification training. The class-based semantic information is conveyed to the sample regions in the R-MAC block. Thus the different regions responding to the classification devote their varying contribution to generating the R-MAC descriptor. The learned R-MAC descriptor containing class-based semantic information can help address the \textbf{SPDD} problem by differentiating regions with a similar texture. Put an example of a chest X-ray, although two chest nodules with different sizes are the same texture, they are labeled benign and malignant and diagnosed with different diseases, respectively. As shown in Fig. \ref{fig3}, the two chest nodules that belonged to different classes vary differently in the learned feature maps of the convolutional layer. By training the R-MAC branch, the R-MAC descriptor can exploit the class-based semantic information of chest nodules and feedback to the convolutional layer in the R-MAC block, then the core node of the main branch.

\subsection{FPN Branch}
Pixel-wise segmentation help extract features that emphasize the pathological abnormal regions. Beneficial from the segmentation training, the FPN branch explores the multi-scale subtle differences of pathological regions at different stages and then give feedback to the core node in the main branch. FPN leverages the convolutional features from low to high levels to extract multi-scale spatial information by building a pyramidal feature hierarchy. FPN has been a criteria component in the network of object detection and shows its powerful feature extraction capability to achieve higher accuracy \cite{liu2016ssd,he2017mask}. The multi-scale spatial information of pathological regions helps generate differentiating features in medical instance retrieval. In our Y-Net, we leverage the FPN components to extract multi-scale spatial information from medical images for semantic segmentation by setting the label of mask images semantically. The structure of the FPN branch is introduced as follows:
\begin{itemize}
    \item Following the core node in the main branch, the FPN branch provides two convolutional layers and three top-down blocks. Last, the FPN branch generates a predicted mask image for pixel-wise segmentation training. The segmentation loss is feedback to the FPN branch and the main branch to help exploit the multi-scale subtle differences of pathological regions at different stages.
    \item In the feed-forwarding process, corresponding to the three bottom-up blocks \{$32\times 64\times 64$ as b1, $64\times 32\times 32$ as b2, $128\times 16\times 16$ as b3\} in the main branch, three top-down blocks \{$32\times 16\times 16$ as t3, $32\times 32\times 32$ as t2, $32\times 64\times 64$ as t1\} in the FPN branch merge them by element-wise addition. The outputs of two bottom-up blocks \{b2, b3\} convolute into $32$-dimensions channel. The output of the convolutional layer before block \{t3\} and two top-down blocks \{t2, t3\} are resized into twice times their width and height by bi-linear up-sampling. Then, the convolutional layer before top-down block \{t3\} and the bottom-up block \{b3\}, the top-down block \{t3\} and the bottom-up block \{b2\}, the top-down block \{t2\} and the bottom-up block \{b1\}, these pairs with the same spatial size are merged by element-wise summation. The addition operation of these three pairs generate the top-down blocks \{t3, t2, t1\} successively. Last, the top-down block \{t3\} convolutes into the size of the mask image.
\end{itemize}

\begin{figure}[htbp]
  \centering
  \includegraphics[width=0.95\linewidth]{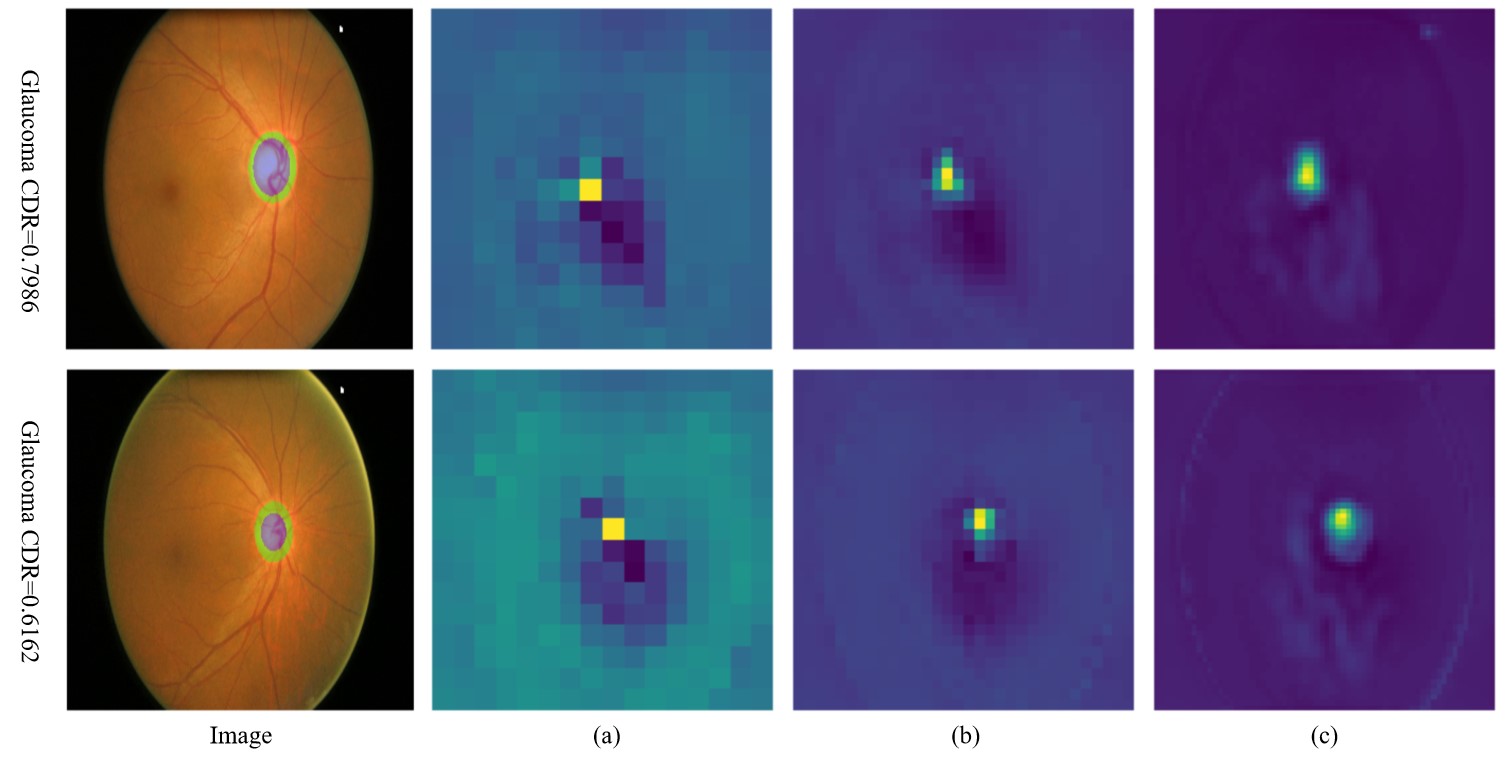}
  \caption{Feature maps of the FPN branch. We calculate the mean value along the channel axis of the features of the three top-down blocks \{t3, t2,t1\}, then visualize the mean value: (a) the feature map of the top-down block \{t3\} with $16\times 16$, (b) the feature map of the top-down block \{t2\} with $32\times 32$, and (c) the feature map of the top-down block \{t1\} with $64\times 64$. }
  \label{fig4}
\end{figure}

In the main branch, the features of bottom-up blocks with lower-level information are more accurately localized by sub-sampling. In the FPN branch, the features of top-down blocks with higher-level information have a stronger spatial resolution by up-sampling. The features of top-down blocks can be enhanced by merging the features from bottom-up blocks. Based on the segmentation training, the learned pyramidal features can learn multi-scale spatial information and are feedback to the main branch's core node. As shown in Fig. \ref{fig4}, the subtle CDR differences between two glaucoma images can be observed on the learned feature maps. The minor difference reflects the different information on different CDRs of same glaucoma, and the difference is encoded into the core node. Based on the pixel-wise segmentation training, we argue that the multi-scale subtle differences of pathological regions at different stages can be learned by the FPN branch to tackle the \textbf{DPSD} problem. 

\subsection{Coupled Loss}
In Y-Net, we integrate the classification task in the R-MAC branch and the segmentation task in the FPN branch to learn the semantic and spatial information of pathological regions simultaneously. To balance the two tasks' loss, we design a coupled loss to unify the classification and segmentation learning. In general, the gradient size is different in the convergence process of different tasks, and the sensitivity to different learning rates is also different. Unifying the scale of different loss functions can prevent the loss items with small gradients from being covered by the loss items with large gradients. Unifying the losses to the same order of magnitude can help improve the generalization of the learned features \cite{kendall2018multi}. The coupled loss function is defined as:
\begin{equation}
    \begin{aligned} 
    \mathcal{L} = \omega \mathcal{L}_{l} + (1-\omega) \mathcal{L}_{r}
    \end{aligned}
    \label{eq2}, 
\end{equation}
where $\mathcal{L}_{r}$ is the circle loss \cite{sun2020circle} for the classification training, $\mathcal{L}_{l}$ denotes the cross-entropy (CE) loss for the pixel-wise segmentation training, and $\omega$ is the weight factor. In the circle loss, each similarity score is given different penalties according to its distance to the optimal effect. In the R-MAC branch, instead of the CE loss, we adopt the circle loss to preserve the class-aware similarity of pathological regions and help prevent minimizing intra-class distance. 

\begin{figure}[htbp]
  \centering
  \includegraphics[width=0.95\linewidth]{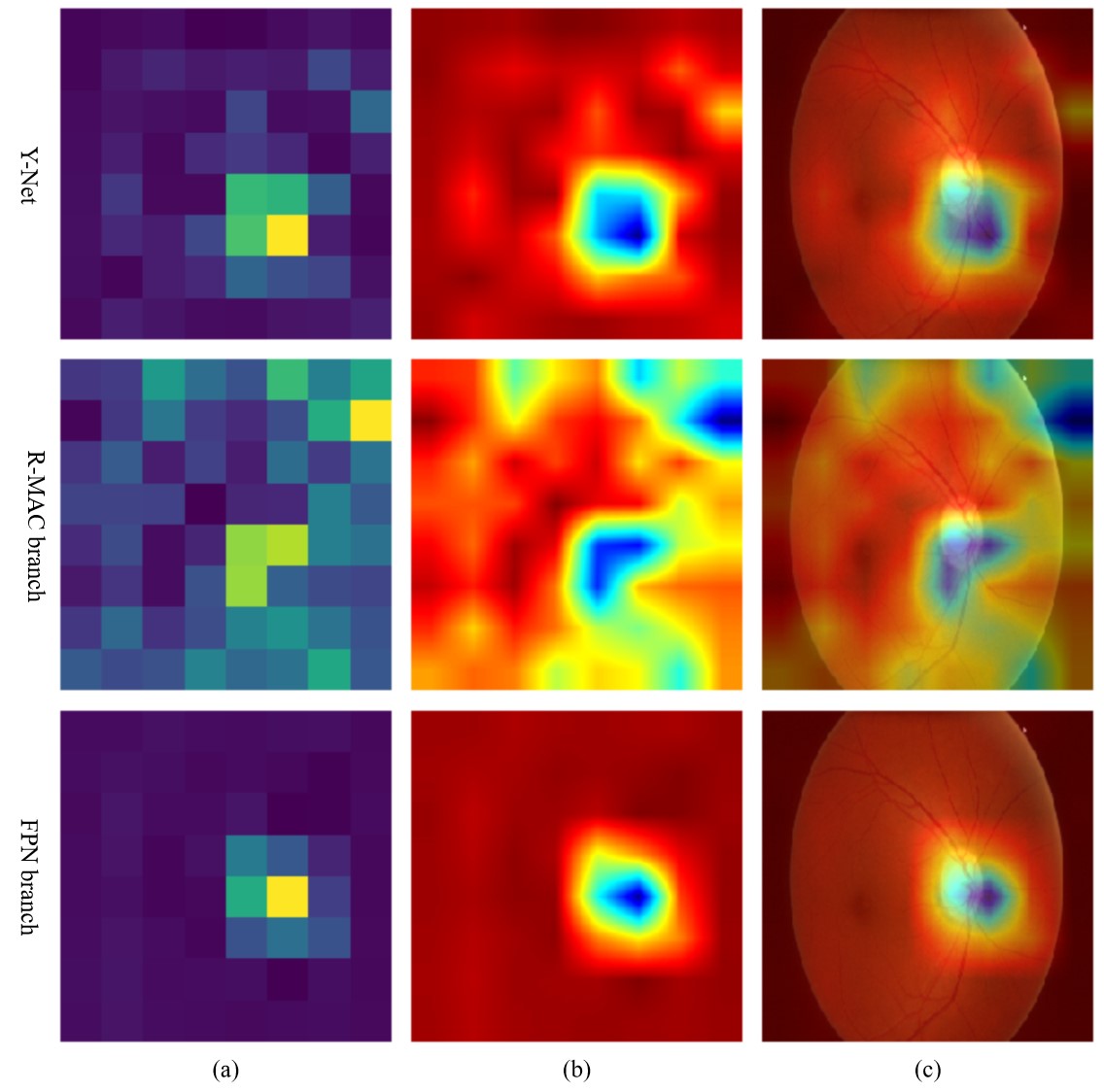}
  \caption{Feature maps of the core node in the main branch. We calculate the mean value along the channel axis of the convolutional features with $256\times 8\times 8$, then visualize the mean value: (a) the feature map with $8\times 8$, (b) the color map resized to the size of the input image with $256\times 256$, and (c) the overlay map combined the color map with the input image.}
  \label{fig5}
\end{figure}

Based on the coupled loss unifying the classification loss and segmentation loss, the main branch's core node can effectively retain the multi-scale spatial information from segmentation training and the class-aware semantic information from classification training simultaneously. The convolutional feature maps from a certain convolutional layer can be viewed as an array of local features sampled from a dense sampling grid. In Fig. \ref{fig5}, the pathological region is the cup and disk of glaucoma. By observing the core node's feature maps, the FPN branch focuses on exploring the pathological region (cup and disk), and the R-MAC branch concerns the highly activated region of the whole image (glaucoma). With the help of the coupled loss balancing the two losses, the Y-Net row's feature maps confirm the effectiveness of preserving the information from the R-MAC branch and the FPN branch. Hence, the learned convolutional features of the core node can be used to generate hash-codes to combat pathological regions' ambiguous manifestations. 

\section{Experiments and Analysis}
To evaluate the performance of our proposed Y-Net, we conduct extensive experiments on two public medical image datasets to verify our method's effectiveness in combating the ambiguous manifestation of pathological regions. In this section, we will introduce the experimental details and analyze the experimental results. 

\subsection{Datasets}
\textbf{Fundus} \cite{cheng2017similarity} contains 650 annotated retina images. Each image is tagged with classification information and manually segmented the result of optic disc and cup. This dataset is obtained from a population-based study and is therefore suitable for evaluating glaucoma screening performance. In this dataset, 168 images from glaucomatous eyes and 482 images from normal eyes are classified. Manual CDR computed from the manually segmented disc, and cup boundaries are necessary for segmentation training. Based on the classification and segmentation labels, we split this dataset into the train set and the test set by ratio $9:1$. The test set of 65 consists of 16 glaucoma images and 49 normal images, and the train set of 585 images covers 152 glaucoma images and 433 normal images. 

\textbf{JSRT} \cite{shiraishi2000development} provides 154 nodule and 93 non-nodule chest X-ray images. Each nodule case contains a nodule only, which is rated as benign or malignant by 20 different radiologists. A detailed delineation of the segmentation's nodule is publicly available to train a lung segmentation \cite{van2006segmentation}. This dataset annotates the lesion position and responding diagnosis. For example, the lesion region of a malignant image is located on the left lung's upper lobe and diagnosed as lung cancer. The annotation images for segmentation tasks are binary images in which pixels are either 255 for the foreground or 0 for the background. We sample 138 images containing 89 malignant nodules and 49 benign nodules to form a train set and 16 images containing 11 malignant nodules and 5 benign nodules to form a test set. The ratio of the train set and the test set is $9:1$.

\subsection{Experimental Setups}
We mainly use mean average precision (\textbf{mAP}) for quantitative evaluation. In the returned list, mAP averages the ranks of images similar to the query image to measure the rank quality. The mAP is usually adopted for evaluating the retrieval performance \cite{zhou2017recent,zheng2017sift}, and is calculated as follows:
\begin{equation}
    \begin{aligned} 
        AP = \frac{\sum_{k=1}^{n} P(k)\cdot rel(k)}{R} 
    \end{aligned}
    \label{eq3}, 
\end{equation}
where $R$ denotes the number of similar results for the current query image, $P(k)$ denotes the precision of top-$k$ retrieval results, $rel_{k}$ is a binary indicator function equaling 1 when the $k$-th retrieved results is similar to the current query image and 0 otherwise, and $n$ denotes the total number of retrieved results. Based on the class labels and the aim of instance retrieval assisting the clinician's own decision-making by reviewing similar cases, the success criteria of similar images are defined as that the two images have similar pathological patterns.

Y-Net is compared against several representative approaches of instance-level retrieval. The comparative approaches are categorized as: weight feature aggregating on convolutional features, regional feature aggregating on convolutional features, and feature embedding from a full-connected layer. 
\begin{itemize}
    \item \textbf{Weight feature aggregating.} \textbf{CroW} \cite{kalantidis2016cross} estimates a spatial weighting of the features as a combination of convolutional feature maps across all channels of the layer. Features at locations with salient visual content are boosted while weights in non-salient locations are decreased. To explicitly leverage semantic information, \textbf{CAM} \cite{jimenez2017class} obtains semantic-aware weights for convolutional features by exploiting the predicted classes. CAM generates a set of spatial maps highlighting the contribution of the regions within an image. Each map is used to weigh the convolutional features and generate a set of class vectors that are aggregated as the region vectors over the fixed region strategy of R-MAC. CAM inspires our R-MAC branch of Y-Net. \textbf{BLCF} \cite{mohedano2018saliency} builds an efficient image representation by combining saliency weighting over convolutional features aggregated by using a large vocabulary with a bag of words (BoW) model \cite{brownlee2017gentle}. \textbf{SOLAR-Local} \cite{ng2020solar} focuses on second-order spatial information to learn local patch descriptors without extra supervision. Based on the feature weighting strategy \cite{revaud2019learning}, it combines the second-order spatial attention and the second-order descriptor loss to improve image features for retrieval and matching.
    \item \textbf{Regional feature aggregating.} \textbf{R-MAC} \cite{tolias2015particular} is an aggregation method for convolutional features to generate a set of regional vectors by performing spatial max-pooling within a particular region. Building on the R-MAC descriptor, \textbf{R-MAC + RPN} \cite{gordo2017end,revaud2019learning} can enhance the ability to focus on relevant regions in the image by replacing the rigid grid with a region proposal network (RPN) trained to localize regions of interest in images. \textbf{Regional Attention} \cite{kim2018regional} presents a context-aware regional attention network for tackling the problem of region-based feature aggregation suffering from the background clutter and varying importance of regions, especially in R-MAC, by weighting an attentive score of a region. \textbf{Deep Vision + SOLO} \cite{salvador2016faster} is trained for instance-level retrieval of image- and region-wise representations pooled from an object detection CNN. In this experiment, we take advantage of the object proposals learned by SOLO \cite{wang2019solo} and their associated convolutional features to build an instance search pipeline. 
    \item \textbf{Feature embedding.} Three deep hashing methods using feature embedding are used to build benchmarks for our Y-Net, including \textbf{DPSH} \cite{li2015feature}, \textbf{DSH} \cite{wang2016deep}, \textbf{DRH} \cite{conjeti2017hashing}, \textbf{DDMH} \cite{yang2020deep}. DPSH performs simultaneous feature learning and hash-code learning with deep neural networks by maximizing pairwise similarities. Inspired by DPSH, DSH proposes a triplet label-based deep hashing method to maximize the given triplet labels' likelihood. DRH offers good separability of classes in hashing space while preserving semantic similarities in local embedding neighborhoods for supervised hashing of medical images through residual learning. DPSH and DSH use AlexNet \cite{krizhevsky2012imagenet} as the backbone. Recently, the residual block \cite{he2016deep} has been used popularly as the backbone in deep hashing methods such as DRH and shows the advantage of feature extraction. In our Y-Net, the main branch also uses the residual block as the backbone. DDMH proposes a unique disentangled triplet loss to effectively push positive and negative sample pairs by desired Hamming distance discrepancies for hash-codes with different lengths.
\end{itemize}

Our Y-Net is implemented under the PyTorch framework, and experiments are run on Geforce RTX 2080 Ti. In our work, the indexing and similarity calculation for evaluation uses Faiss \cite{johnson2019billion}, a library for efficient similarity search and clustering of dense vectors. We use the mini-batch stochastic gradient descent with $0.9$ momentum. The mini-batch size of images is fixed as $32$, and the weight decay parameter is $0.001$. All deep models are trained from scratch with $500$ epochs. It spends approximately $3$ hours for training our Y-Net. The pixel-wise cross-entropy loss is used in the segmentation task. The circle loss \cite{sun2020circle} is used for classification training by using cosine similarity and setting a scale of $32$, a margin of $0.25$. The weight factor $\omega$ in the coupled loss is initially set as $0.5$. We use the $5$-fold cross-validation to select the best classification and segmentation model. The parameters of comparative methods are set according to their implementation details in the corresponding papers, and the best performance is reported. Based on top-$10$ retrieval results, we investigate our Y-Net's performance over hash-code with lengths of $36$, $64$, $128$, $256$, respectively. According to Table \ref{tb1}, with the hash-code lengthen, the performance can correspondingly improve at the cost of storage and search efficiency. As a trade-off between performance and search cost, we report all the performances over $64$-bits hash-code for our Y-Net. 
\begin{table}[!t]
\renewcommand{\arraystretch}{1.0}
\caption{mAP of Y-Net over the varying length of hash-codes on the Fundus and JSRT datasets.}
\begin{center}
    \begin{tabular}{|c|c|c|c|c|}
    \hline 
    \textbf{Datasets} & \textbf{mAP@36} & \textbf{mAP@64} & \textbf{mAP@128} & \textbf{mAP@256} \\
    \hline
    \textbf{Fundus} & 0.5903 & 0.6102 & 0.6266 & 0.6308  \\
    \textbf{JSRT} & 0.5361 & 0.5518 & 0.5732 & 0.5809  \\
    \hline
    \end{tabular}
\end{center}\label{tb1}
\end{table}

\subsection{Experimental Results}
The following research questions will be answered by analyzing experimental results: 
\begin{description}
\item[\textbf{RQ1}] Does our proposed Y-Net outperform the state-of-the-art methods on retrieval performance in medical instance retrieval? 
\item[\textbf{RQ2}] Can our proposed Y-Net help to combat the ambiguity of pathological regions in medical instance retrieval?
\item[\textbf{RQ3}] What are the effectiveness of the R-MAC branch, the FPN branch, and the coupled loss in our proposed Y-Net framework?
\item[\textbf{RQ4}] How is the retrieval efficiency of our proposed Y-Net?
\end{description}

\subsubsection{Quantitative Analysis (RQ1)}
\begin{table*}[!t]
\renewcommand{\arraystretch}{1.0}
\caption{mAP over the varying number of the returned list on the Fundus and JSRT datasets.}
\begin{center}
    \begin{tabular}{|c|c|c|c|c|c|c|c|c|c|}
    \hline 
    \multirow{2}{*}{\textbf{Methods}} & \multirow{2}{*}{\textbf{Dim}} & \multicolumn{4}{c|}{\textbf{Fundus}} & \multicolumn{4}{c|}{\textbf{JSRT}}\\
    \cline{3-10}
     && \textbf{top-5} & \textbf{top-10} & \textbf{top-20} & \textbf{top-50} & 
        \textbf{top-5} & \textbf{top-10} & \textbf{top-20} & \textbf{top-50} \\
    \hline
    \textbf{CroW \cite{kalantidis2016cross}} & 512 & 0.5223 & 0.4681 & 0.4471 & 0.4366 & 0.4993 & 0.4705 & 0.4396 & 0.4189 \\
    \textbf{CAM \cite{jimenez2017class}} & 2048 & \underline{0.5917} & \underline{0.5488} & 0.4982 & 0.4609 & \underline{0.5611} & \underline{0.5124} & 0.4497 & 0.4187 \\
    \textbf{BLCF \cite{mohedano2018saliency}} & 1000 & 0.4890 & 0.4793 & 0.4463 & 0.4216 & 0.4701 & 0.4356 & 0.4096 & 0.3903 \\
    \textbf{SOLAR-Local \cite{ng2020solar}} & 1024 & 0.5701 & 0.5274 & 0.4766 & 0.4482 & 0.5443 & 0.4987 & 0.4264 & 0.4051\\
    \hline
    \textbf{R-MAC \cite{tolias2015particular}} & 512 & 0.5016 & 0.4884 & 0.4585 & 0.4528 & 0.4682 & 0.4191 & 0.3965 & 0.3812 \\
    \textbf{R-MAC + RPN \cite{revaud2019learning}} & 3072 & 0.5483 & 0.5024 & 0.4685 & 0.4446 & 0.4805 & 0.4461 & 0.4098 & 0.3951 \\
    \textbf{Regional Attention \cite{kim2018regional}} & 2048 & 0.5674 & 0.5279 & 0.5070 & 0.4854 & 0.4984 & 0.4621 & 0.4289 & 0.4069 \\
    \textbf{Deep Vision + SOLO \cite{salvador2016faster}} & 3072 & 0.5486 & 0.5001 & 0.4889 & 0.4815 & 0.5123 & 0.4756 & 0.4358 & 0.4123 \\
    \hline
    \textbf{DPSH \cite{li2015feature}} & 64 & 0.5044 & 0.4693 & 0.4451 & 0.4270 & 0.4581 & 0.4203 & 0.3891 & 0.3677 \\
    \textbf{DSH \cite{wang2016deep}} & 64 & 0.5052 & 0.4882 & 0.4788 & 0.4734 & 0.5487 & 0.4921 & 0.4578 & 0.4332 \\
    \textbf{DRH \cite{conjeti2017hashing}} & 64 & 0.5712 & 0.5435 & \underline{0.5322} & \underline{0.5203} & 0.5306 & 0.4912 & \underline{0.4651} & \underline{0.4498} \\
    \textbf{DDMH \cite{yang2020deep}} & 32 & 0.5231 & 0.5051 & 0.4962 & 0.4802 & 0.5396 & 0.4869 & 0.4421 & 0.4284\\
    \hline
    \textbf{Y-Net (ours)} & 64 & \textbf{0.6367} & \textbf{0.6102} & \textbf{0.5820} & \textbf{0.5581} & \textbf{0.6013} & \textbf{0.5518} & \textbf{0.5284} & \textbf{0.4976} \\
    \hline
    \end{tabular}
\end{center}\label{tb2}
\end{table*}
The performance of the mAP over the returned list of $5$, $10$, $20$, and $50$ on Fundus and JSRT datasets are reported in Table \ref{tb2}, respectively. On the whole, when the returned list lengthens, all methods' performance declines to some extent. Our Y-Net all achieves significant gains of mAP over the varying returned list on the two datasets. Experimental results on the Fundus dataset show that Y-Net outperforms the second-highest methods (underline) by 7.60\%, 11.18\%, 9.35\%, 7.26\% correspond to the different number of the returned list. Y-Net also achieves the best performance on the JSRT dataset compared to the other methods. For the methods obtaining the second-highest performance, CAM is a weighing feature method aggregating on convolutional features, and DRH is a method of feature embedding. This demonstrates that the methods of regional feature aggregating on convolutional features may lose related information between regions after region proposals. This loss prevents them from obtaining better performance. Among methods of weight feature aggregating, SOLAR-Local yields good performance by exploiting the second-order spatial information. CAM can achieve better performance than SOLAR-Local by exploiting class semantic information. The retrieval performance on the Fundus dataset is higher than that on the JSRT dataset by $10.58\%$ on the returned list of $10$. The reason for this gap has two points. The shortage of specificity is the main challenge for chest X-ray image analysis tasks. The JSRT dataset only provides lung masks but not lesion masks; those non-lesion regions in the lung mask may affect the discriminative information learning.

Compared to DRH, CAM acquires a better performance over the returned list of $5$ and $10$. This demonstrates that it effectively explore pathological regions and weigh their activations by exploiting the correlation between class labels and pathological regions. Inspired to CAM, the R-MAC branch in our Y-Net contributes to increasing the retrieval performance by focusing on the pathological regions and weights these regions with class activations. Benefiting from adopting the residual block as the backbone, DRH is superior to CAM over the returned list of $20$ and $50$. To further improve the performance over the longer returned list, we need to exploit spatially subtle differences of pathologically abnormal regions with the help of pixel-wise segmentation training in the FPN branch. Due to the differentiating ability of the subtle differences in pathological regions, our Y-Net surpasses DRH over the returned list of $20$ and $50$ compared to CAM. In summary, three points contribute to the performance of our Y-Net. (1) The R-MAC branch learns the class-aware semantic information of pathological regions. (2) The FPN branch explores the multi-scale subtle spatial information of pathological regions. (3) The main branch uses the residual block as the backbone.

\subsubsection{Qualitative Analysis (RQ2)}
\begin{figure*}[htbp]
  \centering
  \includegraphics[width=0.95\linewidth]{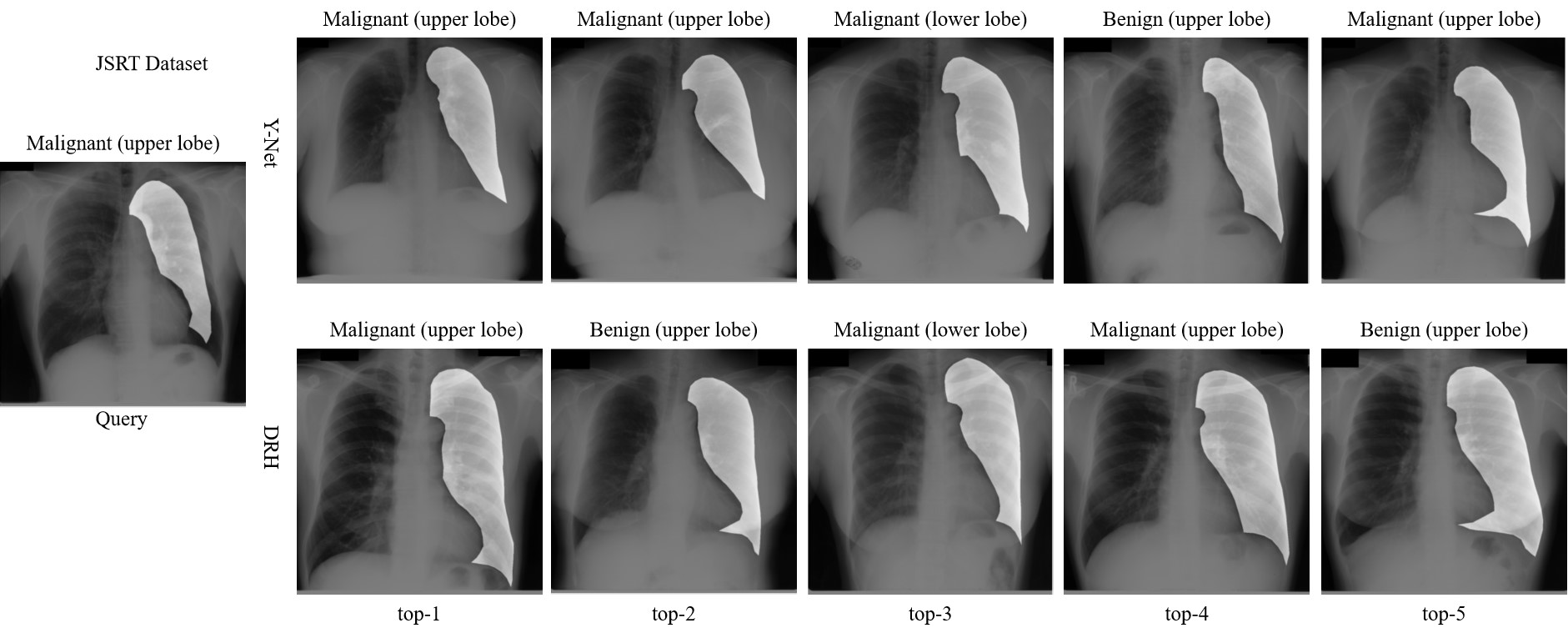}
  \caption{Ranking of the top-$5$ returned list on the JSRT dataset. We query a malignant image and obtain the ranking of the top-$5$ returned list for Y-Net and DRH, respectively. Each image shows the position of the chest nodule labeled manually.}
  \label{fig6}
\end{figure*}

Lung nodules are small masses of tissue in the lung and quite common. They appear as round, white shadows on a chest X-ray. Lung nodules are usually about $0.2$ inches ($5$ millimeters) to $1.2$ inches ($30$ millimeters) in size. A larger lung nodule, such as $30$ millimeters or larger, is more likely to be cancerous than a smaller lung nodule. The regions of chest nodules in X-ray images are hard to differentiate malignant or benign according to the spatial information, including texture and size. So this is a typical \textbf{SPDD} problem. As shown in Fig. \ref{fig6}, our Y-Net returns more malignant images and ranking ahead than DRH by querying a malignant image. Based on the FPN branch exploiting spatially subtle differences of nodule regions, the R-MAC branch cooperatively encodes the class-aware semantic information of pathological regions into the hash-codes. By exploiting the correlation between class labels and pathological regions, the R-MAC branch can address the \textbf{SPDD} problem in medical instance retrieval. In fact, the R-MAC branch weighs the regional of maximum activation by conveying the class-based semantic information to the R-MAC descriptor and the convolutional features. The class-weighted regional of maximum activation can differentiate the same performance of different diseases of medical images.

\begin{figure*}[htbp]
  \centering
  \includegraphics[width=0.95\linewidth]{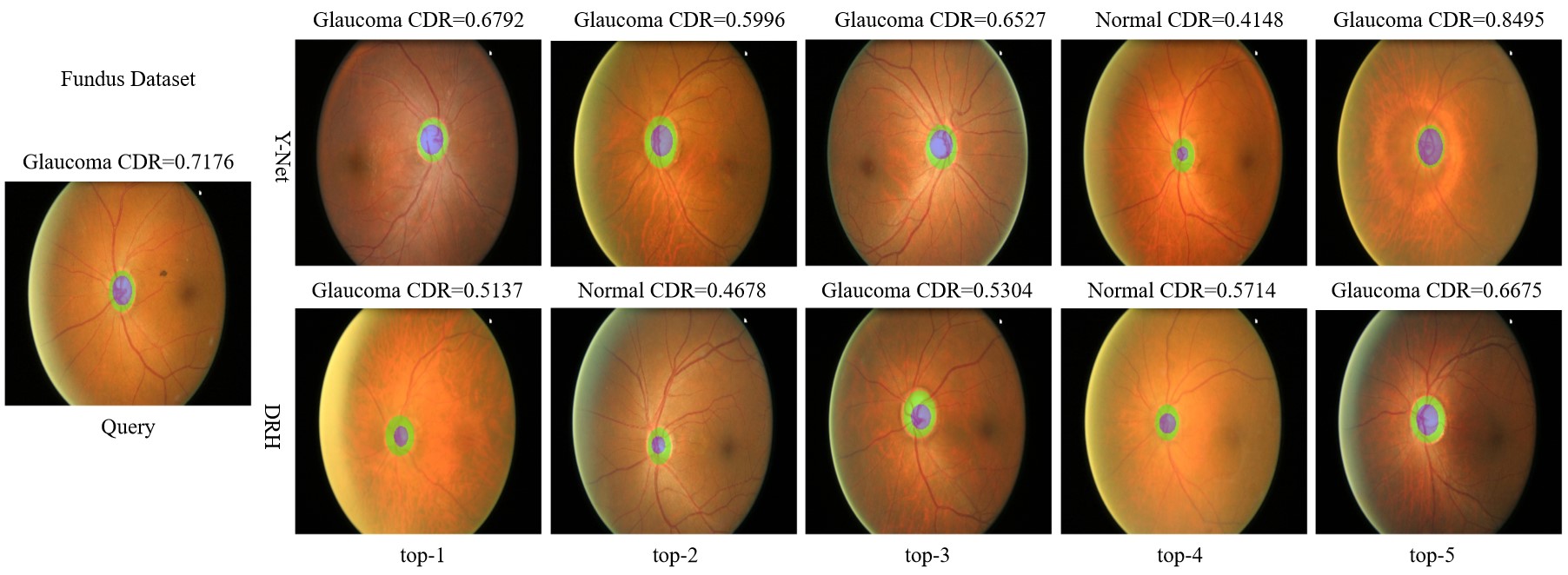}
  \caption{Ranking of the top-$5$ returned list on the Fundus dataset. We query a glaucoma image and obtain the ranking of the top-$5$ returned list for Y-Net and DRH, respectively. Each image shows its Cup to Disk Ratio (CDR) size labeled manually.}
  \label{fig7}
\end{figure*}
The size of CDR computation from color fundus images is the main clue for glaucoma diagnosis \cite{li2021applications}. The different size of CDR denotes different grading of glaucoma. It is useful for clinicians to find the most similar images with closer CDR sizes to make a medico-decision. As shown in Fig. \ref{fig7}, compared to the DRH, Y-Net returned more glaucoma images with closer CDR sizes and ranked ahead by querying a glaucoma image. According to this experimental result, we argue that the FPN branch can effectively encode the subtle differences of pathological regions into the hash-codes to address the \textbf{DPSD} problem by mining the multi-scale spatial information. The FPN branch can locate pathological regions' subtle differences at different stages of the same disease based on the pixel-wise segmentation training. In essence, the FPN branch weights the pathological regions by segmentation training. The weighted pathological regions can be encoded as the most discriminative parts of the hash-codes to differentiate the same disease's different manifestations at different stages. 

Our Y-Net's R-MAC branch exploits the class semantic information to weigh regions of maximum activation to tackle the SPDD problem. Apart from the same pathological criteria evaluation (benign and malignant), we also apply the disease label to evaluate the performance to embody the effectiveness of tackling the SPDD problem. The large disease label consists of lung cancer, granuloma, cryptococcosis, inflammatory mass, etc. The fine disease label for lung cancer includes adenocarcinoma, large cell carcinoma, small cell carcinoma, etc. On the returned list of $10$, our method outperforms CAM by $8.12\%$ average precision on diagnosing disease. This demonstrates that our method can effectively differentiate the similar manifestation of different diseases. Our Y-Net's FPN branch explores the spatially subtle differences of the lesion region to overcome the DPSD problem. Regarding the DPSD problem, we apply average CDR to evaluate the performance on differentiating the different manifestations of the same disease in different stages. Our Y-Net yields the average CDR gap of $0.2157$ between the query image and the retrieved images, while CAM obtains $0.3521$. The convolutional features in the core node of the main branch learn the information from both branches to promote hash-codes' discriminative ability.

\subsubsection{Ablation Study (RQ3)}
\begin{table*}[!t]
\renewcommand{\arraystretch}{1.0}
\caption{mAP of branches of Y-Net over the varying number of the returned list on the Fundus and JSRT datasets.}
\begin{center}
    \begin{tabular}{|c|c|c|c|c|c|c|c|c|}
    \hline 
    \multirow{2}{*}{\textbf{Branches}}  & \multicolumn{4}{c|}{\textbf{Fundus}} & \multicolumn{4}{c|}{\textbf{JSRT}}\\
    \cline{2-9}
     & \textbf{top-5} & \textbf{top-10} & \textbf{top-20} & \textbf{top-50} & 
       \textbf{top-5} & \textbf{top-10} & \textbf{top-20} & \textbf{top-50} \\
    \hline
    \textbf{Y-Net w/o FPN and R-MAC branch}  & 0.5001 & 0.4871 & 0.4679 & 0.4575 & 0.5324 & 0.4856 & 0.4509 & 0.4297 \\
    \textbf{Y-Net w/o FPN branch}  & 0.5881 & 0.5656 & 0.5443 & 0.5033 & 0.5325 & 0.5114 & 0.4831 & 0.4501 \\
    \textbf{Y-Net w/o R-MAC branch}  & 0.5561 & 0.5179 & 0.4854 & 0.4536 & 0.5210 & 0.4914 & 0.4597 & 0.4285 \\
    \textbf{Y-Net w/o Circle loss}  & 0.6061 & 0.5879 & 0.5554 & 0.5136 & 0.5684 & 0.5291 & 0.4976 & 0.4703 \\
    \textbf{Y-Net} & \textbf{0.6367} & \textbf{0.6102} & \textbf{0.5820}  & \textbf{0.5581} & \textbf{0.6013} & \textbf{0.5518} & \textbf{0.5284} & \textbf{0.4976} \\
    \hline
    \end{tabular}
\end{center}\label{tb3}
\end{table*}
To further research the R-MAC branch and FPN branch's contribution, we conduct an ablation study by cropping the corresponding branch of Y-Net. As shown in Table \ref{tb3}, Y-Net without the FPN branch can achieve better performance than Y-Net without the R-MAC branch, and Y-Net achieves convincing performance by unifying the FPN branch and R-MAC branch. Without the FPN branch, Y-net can achieve competitive performance compared to CAM and DRH. Upon the R-MAC branch, Y-Net can obtain a significant gain by adding the FPN branch. This demonstrates that the R-MAC branch can differentiate pathological regions' similar manifestations by weighing the regional of maximum activation based on the classification training. The added gain benefits from the FPN branch, which exploits the subtle differences of pathological regions by mining the multi-scale spatial information based on the segmentation training. As shown in Fig. \ref{fig7}, the glaucoma images ranked ahead are closer to the query image in CDR size. This also confirms the FPN branch's effectiveness in preventing the R-MAC branch from minimizing the intra-class distance. Based on this joint learning scheme, the core node in the main branch absorbs the class-aware semantic information from the R-MAC branch and spatially subtle differences from the FPN branch, then are mapped into the hash-codes. The learned hash-codes can be used to combat the ambiguous manifestations of pathological regions.  

\begin{figure}[htbp]
  \centering
  \includegraphics[width=0.95\linewidth]{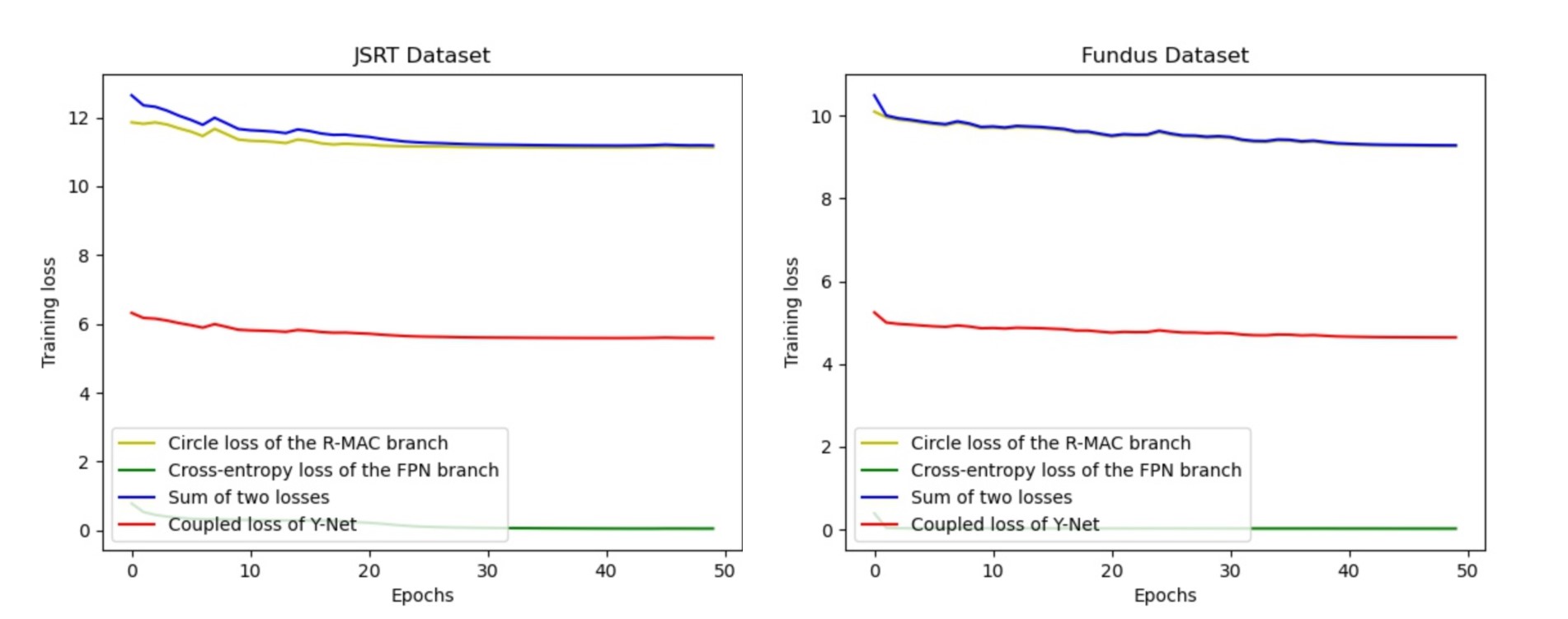}
  \caption{Qualitative results of the coupled loss. The different loss of top-$50$ iterations during training on the Fundus and JSRT datasets are compared.}
  \label{fig8}
\end{figure}
Based on the above experimental analysis, we confirm the effect of unifying classification and segmentation. Next, we would like to discuss the coupled loss function's effect in both branches' unified training. First, as shown in Table \ref{tb3}, the R-MAC branch with the circle loss achieves better performance than the R-MAC branch with the CE loss. The circle loss can help the R-MAC branch maximize the intra-class similarity and minimize inter-class similarity by pair similarity optimization. Second, Compared to the sum of the two losses, the coupled loss function can improve retrieval performance averagely by 2\% on mAP over the varying number of the returned list on the Fundus and JSRT datasets. This demonstrates that the coupled loss can help facilitate the generalization of the learned convolutional features by unifying the losses to the same order of magnitude. As shown in Fig \ref{fig8}, compared to the sum of the two losses (blue), the coupled loss (red) unifies the scale of the circle loss (yellow) and the cross-entropy loss (green) to prevent the loss unbalance in the convergence process of different tasks. In the process of screening and diagnosis, the ambiguous manifestation of pathological regions may be varied. Hence, the two tasks can be mutually beneficial to enhance Y-Net's generalization by the joint learning scheme. 

\subsubsection{Retrieval Efficiency Analysis (RQ4)}
In this section, we discuss the efficiency of the proposed Y-Net from three-folds by putting the Fundus dataset as an example.
\begin{itemize}
 \item[1)] \textbf{Feature computation time.} Based on the pre-trained Y-Net model, we inference the hash-codes of $64$-bits from the core node in the main branch. Hence, the feature computation of the main branch occupies the most time cost in the test stage. We can complete the hash-codes generating for the training set of $585$ images in $4$ seconds on GPU. The feature computation time of our Y-Net is fair to the most comparative methods.
 \item[2)] \textbf{Retrieval time.} After hash-codes generating, we build the index in $1$ second for the training set by using Faiss. By querying the test set of $65$ images, returning top-$10$ most similar images can be done in $34 ms$. The time-consuming processes of the search engine are the indexing search and similarity calculation. The time cost of both lengthens when the size of feature vectors used for similarity calculation extends. As Table \ref{tb2} shows (column: Dim), Y-Net's hash-code length is equal to the methods using feature embedding.
 \item[3)] \textbf{Memory cost.} The memory-consuming is about $2,000$ Mbps during model training by setting the batch size at $32$. The online search for the index also consumes about $2,000$ Mbps. The memory cost depends on the model complexity where our Y-Net is fair to the methods aggregating regional features.
\end{itemize}
According to the above analysis of efficiency, our Y-Net can provide fair real-time responses with significantly improving the performance by comparing to the state-of-the-art methods.

\section{Conclusions}
To combat the manifestation ambiguity in medical instance retrieval, we propose a novel framework called Y-Net, encoding images into compact hash-codes aggregating from convolutional features. The proposed Y-Net contains the main branch, the R-MAC branch, the FPN branch. Based on the classification loss, the R-MAC branch encodes the class-aware semantic information of pathological regions into the convolutional features to avoid \textbf{SPDD} problem. And based on the pixel-wise segmentation loss, the FPN branch encodes the spatially subtle differences of pathological regions into the convolutional features to overcome the \textbf{DPSD} problem. After unifying the classification and segmentation training, the learned convolutional features in the main branch are directly aggregated to generate the hash-codes for similarity measure. The extensive experiments on the two medical image datasets with class and pixel-wise mask labels show that our Y-Net can alleviate pathologically abnormal regions' ambiguity. 

There also exist two limitations of this work. First, it is hard to acquire medical image datasets with pixel-wise segmentation annotations, while detecting the subtle differences with the bounding box of pathological regions is challenging. This restricts our Y-Net's availability and universality. Second, the multi-instances and multi-labels of medical images significantly lift the difficulty of combating pathologically abnormal regions' ambiguity. In the future, we would like to explore the solutions to address such issues.

\bibliographystyle{ieeetr} 
\bibliography{ynet}      

\end{document}